\documentclass{article} 
\usepackage{lpiabstr_2e}
\usepackage{times}
\usepackage{epsf}
\usepackage[sectionbib]{natbib}

\begin{document}

\title{The Sun's Dust Disk -- Discovery Potential of the New
Horizons mission during Interplanetary Cruise}

\author{M. Landgraf}
\affil{ESA/ESOC, Robert-Bosch-Strasse 5, 64293 Darmstadt, Germany
(Markus.Landgraf@esa.int)}

\runningtitle{THE SUN'S DUST DISK: M. Landgraf}

\titlemake 

\begin{abstracttext}
\section{Introduction}

When the Pioneer 10 spacecraft entered the interplanetary space beyond
Jupiter's orbit, it detected an almost constant flux of impacts by
dust particles \citep{humes80} larger than $10\:{\rm \mu m}$. This was
unexpected, as the dust from comets, which were the only potential
sources of dust known at the time, is believed to be less concentrated
at larger heliocentric distances. At the time, an exotic distribution
of cometary orbits had to be introduced in order to explain the
Pioneer data. Dust from outside the solar system can not explain the
constant flux detected by the Pioneer experiments, because the
interstellar flux of dust particles large enough to be detectable by
the Pioneer instruments is at least an order of magnitude lower than
the detected flux \citep{landgraf00a}. The discovery of objects in the
Edgeworth-Kuiper belt (EKB) \citep{jewitt93} offered the possibility
for another dust source: The objects in the EKB should produce dust by
mutual collisions and by collisions with interstellar dust particles
\citep{yamamoto98}, forming a disk of dust around the Sun. Modelling the
evolution of the orbits of dust grains from the EKB \citet{landgraf02}
showed, that indeed the Pioneer data can only be explained by dust
migrating in from the EKB under the influence of the
Poynting-Robertson drag.

Unfortunately the Pioneer dust measurements extend only to a
heliocentric distance of $18\:{\rm AU}$, where the instrument ceased
function. Consequently, Pioneer detected merely the inner edge of the
solar system dust disk. Here the question arises, what the radial
density profile of the solar system dust disk is and what processes
govern the evolution of the dust grains after the release from the
parent body. Because we do not know the collisional lifetime of dust
grains from the EKB, it is unclear, what fraction of EKB dust grains
survives the evolution to inside the orbit of Uranus, where Pioneer 10
made its far most measurements. We present a prediction of the dust
detection rate of the planned New Horizons mission using the same
model that was employed to analyse the Pioneer data. We show what the
signature of different dust lifetimes will be in the data collected by
this mission.

\section{Modelling the Solar System Dust Disk} 

The evolution of a dust grain released by an EKB object depends mainly
on two parameters: it's collisional lifetime, and its susceptibility
to solar radiation pressure, which is parameterised by the constant
ratio $\beta$ of radiation pressure force to gravity. Here we assume
values of $\beta$ of $0.03$, $0.08$, $0.2$, and $0.5$. The value of
$\beta$ is mainly controlled by the grain size
\citep{vandehulst57}. Assuming a homogeneous spherical grain with a
bulk mass density of $1\:{\rm g}\:{\rm cm}^{-3}$, the value
$\beta=0.03$ corresponds to a grain diameter of $16\:{\rm \mu m}$,
$\beta=0.08$ to a diameter of $6\:{\rm \mu m}$, $\beta=0.2$ to a
diameter of $3\:{\rm \mu m}$, and $\beta=0.5$ to a diameter of
$1\:{\rm \mu m}$. For the sources of the dust grains we assume all 420
trans-Neptunian objects listed with the Minor Planet Center of the IAU
that have been observed at more than one opposition. Upon release the
dust grains acquire orbits different from those of their parent bodies
due to the effect of solar radiation pressure \citep{kresak76}. 

\begin{figure}[ht]
\centering
\epsfxsize=.8\hsize
\epsfbox{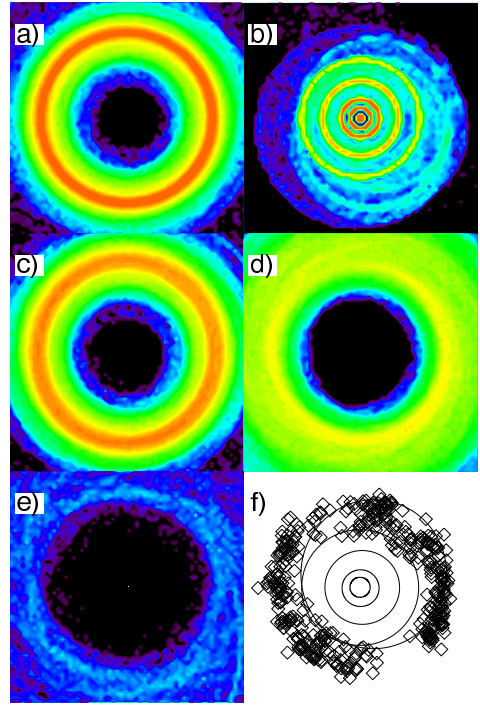}
\caption{\label{fig_discs} Spatial density distribution of dust from
EKB objects (normalised logarithmic colour scale: blue=0.001, red=1) in
a $120\:{\rm AU}\times 120\:{\rm AU}$ sub-plane of the ecliptic. Panels
a), c), d), and e) show the distribution of grains with $\beta=0.03$,
$0.08$, $0.2$, and $0.5$ respectively for a collisional lifetime of
$10^6\:{\rm a}$. In panel b) the distribution for grains with
$\beta=0.03$ is shown if no collisions are considered.  Panel f) shows
the orbits of Jupiter, Saturn, Uranus, Neptune, and Pluto, and the
positions of EKB objects predicted for the epoch of the planned fly-by
of Pluto by the New Horizons mission in 2016.}
\end{figure}

The grain's orbits evolve under the effect of the Poynting-Robertson
drag towards the inner solar system
\citep{liou96b}. On the way inwards they can be trapped in mean motion
resonances (MMRs), or can be destroyed by a collision with an
interstellar or another EKB dust grain. Using the same method as
\citet{liou96b} we have simulated the evolution of grains with
$\beta=0.03$ for a lifetime of $1\times 10^6$ years.  In order to
determine the effects of the frequency of grain destruction by
collisions on the distribution of dust in the disk we simulate grains
with $\beta=0.03$ from one representative source object without
considering collisions. In this case the end of life of the dust
grains is determined by ejection from the solar system by a close
encounter with a giant planet or by falling into the
Sun. Figure~\ref{fig_discs} shows the spatial density distributions
for the different values of $\beta$ and also for the case without
collisions.

The simulation shows that the larger grains (with smaller values of
$\beta$) are concentrated in a ring near the source region. For larger
values of $\beta$ the disk becomes more extended and distributed with
lower peak densities. This is because the grains with higher $\beta$
exhibit high eccentricities and large semi-major axes (typically
$150\:{\rm AU}$ for $\beta=0.5$) immediately after their release. If
there were no collisions even these grains would ultimatively circle
towards the Sun. The case without collisions (figure~\ref{fig_discs}
b)) demonstrates this effect. The inner solar system is well
populated, especially MMRs with the giant planets, which show up as
concentric rings in the figure.

\begin{figure}[ht]
\centering
\epsfxsize=\hsize
\epsfbox{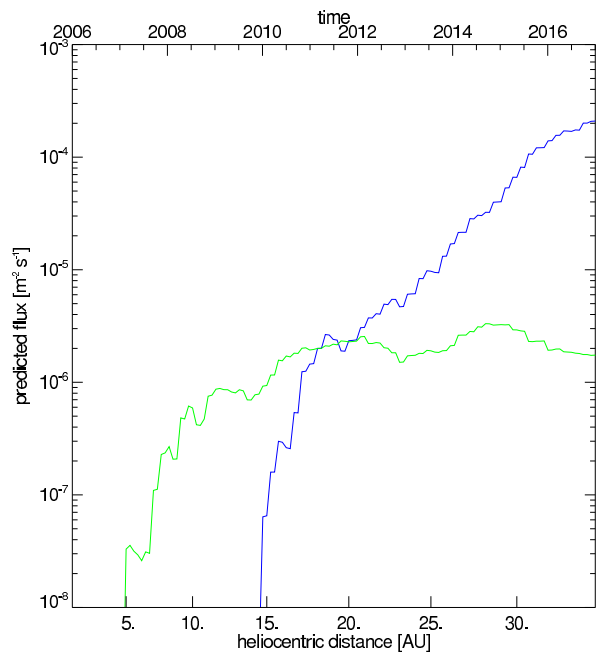}
\caption{\label{fig_pke_fluxes} Prediction of impactor flux
(normalised) onto the dust instrument of the New Horizons mission. The
blue curve shows the prediction for dust grains with $\beta=0.03$ and
a collisional lifetime of $10^6\:{\rm a}$, and the green curve for
grains with $\beta=0.03$ that are not subject to collisional
destruction.}
\end{figure}
\section{Prediction of Impact Flux for New Horizons}
It is planned to launch New Horizons in early 2006 on a
high-energy trajectory towards Jupiter. The fly-by of Jupiter will
send the spacecraft onto an almost radial trajectory towards Pluto,
the fly-by of which will then occur in 2016. Here we assume that the
target area of the dust instrument is mounted on the side of the
spacecraft opposite to the main antenna, which is assumed to point
towards the Sun. The simulation described above allows us to determine
the flux vector of dust from the Kuiper belt relative to the sensitive
target area along the spacecraft's trajectory. The resulting impactor
flux can thus be predicted as a function of time and heliocentric
distance. In order to fix the absolute value of the dust concentration
we scale the prediction so that the flux of grains with $\beta=0.03$
(larger than $16\:{\rm \mu m}$) at $18\:{\rm AU}$ is $2\times
10^-6\:{\rm m}^{-2}\:{\rm s}^{-1}$ as was measured by Pioneer 10. The
prediction shown in figure~\ref{fig_pke_fluxes} shows that the
profiles of the dust flux curves depend strongly on the collisional
lifetimes of the grains. For the case with no collisions (or
collisional lifetime longer than the Poynting-Robertson lifetime) the
flux curve remains relatively flat outside the orbit of Uranus. In the
case where collisions lead to a destruction of the grain after, on
average, $10^6$ years, the predicted flux increases steeply with
heliocentric distance.

\section{Conclusion}
We have simulated the evolution of the distribution of cosmic dust
generated by EKB objects using realistic initial orbits and
considering various values for the strength of radiation pressure and
the collisional lifetime. The simulation together with the measurements of
dust by the Pioneer spacecraft allows a prediction of the expected
impactor flux on the future New Horizons mission.

New Horizons will be able to distinguish two different kinds of dust
evolution in the EKB: driven by collisions or orbit migration under
Poynting-Robertson drag. The main unknown here is the rate of
collisions with interstellar dust particles. If not collisions but
Poynting-Robertson dominate the evolution of EKB dust, the dust flux
measured by New Horizons beyond $20\:{\rm AU}$ can be expected to be
constant at the level that was measured by Pioneer 10. If, on the
other hand, the flux of interstellar dust grains larger than $1\:{\rm
\mu m}$ is $10^{-4}\:{\rm m}^{-2}\:{\rm s}^{-1}$ in the outer solar
system, equal to the flux measured by the Ulysses mission
\citep{gruen94} inside the orbit of Jupiter, then the average
collision rate on a EKB dust grain with a diameter of $16\:{\rm \mu
m}$ is about one every $10^6$ years. The curve in
figure~\ref{fig_pke_fluxes} shows that the dust population detected by
Pioneer 10 at a heliocentric distance of $18\:{\rm AU}$ is then merely
a tiny fraction of the dust that is more abundant by two orders of
magnitude outside $30\:{\rm AU}$, closer to its sources. In that case
New Horizons can expect to find a very high flux of more than
$10^{-4}\:{\rm m}^{-2}\:{\rm s}^{-1}$, or about $10$ hits per day for
each square meter of sensitive area.

\end{abstracttext}


\begin{thebibliography}{}
\markboth{}{THE SUN'S DUST DISK: M. Landgraf}
\setlength{\itemsep}{0pt}
\bibitem[\protect\citename{Gr{\"u}n {\em et~al.\ }\relax, }1994]{gruen94}
{\sc Gr{\"u}n, E., et al.}, 1994. {\em A\&A}, {\bf 286}, 915--924.
\bibitem[\protect\citename{Humes, }1980]{humes80}
{\sc Humes, D.~H.} 1980. {\em JGR}, {\bf 85}(A/II), 5841--5852.
\bibitem[\protect\citename{{Jewitt} \& {Luu}, }1993]{jewitt93}
{\sc {Jewitt}, et al.} 1993. {\em Nature}, {\bf 362}, 730--732.
\bibitem[\protect\citename{Kresak, }1976]{kresak76}
{\sc Kresak, L.} 1976. {\em Bull. Astronom. Inst. Czech.}, {\bf 27}, 35--46.
\bibitem[\protect\citename{Landgraf {\em et~al.\ }\relax, }2000]{landgraf00a}
{\sc Landgraf, M., et al.} 2000. {\em JGR}, {\bf 105}(A5), 10343--10352.
\bibitem[\protect\citename{Landgraf {\em et~al.\ }\relax, }2002]{landgraf02}
{\sc Landgraf, M., et al. } 2002. {\em AJ}, {\bf 123}(5), 2857--2862.
\bibitem[\protect\citename{Liou {\em et~al.\ }\relax, }1996]{liou96b}
{\sc Liou, J.-C., et al.} 1996. {\em Icarus}, {\bf 124}, 429--440.
\bibitem[\protect\citename{van~de Hulst, }1957]{vandehulst57}
{\sc van~de Hulst, H.~C.} 1957.{\em Light scattering by small particles}.
Dover Publications Inc., New York.
\bibitem[\protect\citename{Yamamoto \& Mukai, }1998]{yamamoto98}
{\sc Yamamoto, S., \& Mukai, T.} 1998. {\em A\&A}, {\bf 329}, 785--791.
\end{thebibliography}
\end{document}